\documentclass[a4paper]{article}
\usepackage{INTERSPEECH2022}
\usepackage{multirow}
\title{PHO-LID: A Unified Model Incorporating Acoustic-Phonetic and Phonotactic Information for Language Identification}
\name{Hexin Liu$^1$, Leibny Paola Garcia Perera$^2$, Andy~W.~H.~Khong$^1$, Suzy J. Styles$^3$, Sanjeev Khudanpur$^2$}
\address{
  $^1$School of Electrical and Electronic Engineering, Nanyang Technological University, Singapore\\
  $^2$CLSP and HLT-COE, Johns Hopkins University, USA\\
  $^3$Psychology, School of Social Sciences, Nanyang Technological University, Singapore}
\email{HEXIN002@e.ntu.edu.sg, lgarci27@jhu.edu}

\begin{document}

\abovedisplayskip=4pt
\belowdisplayskip=4pt

\setlength\floatsep{11pt}
\setlength\textfloatsep{11pt}
\setlength\abovecaptionskip{5pt}
\setlength\dblfloatsep{5pt}
\setlength\dbltextfloatsep{11pt}

\maketitle
\begin{abstract}
 We propose a novel model to hierarchically incorporate phoneme and phonotactic information for language identification~(LID) without requiring phoneme annotations for training. In this model, named PHO-LID, a self-supervised phoneme segmentation task and a LID task share a convolutional neural network~(CNN) module, which encodes both {\em language} identity and sequential {\em phonemic information} in the input speech to generate an intermediate sequence of ``phonotactic'' embeddings. These embeddings are then fed into transformer encoder layers for utterance-level LID.  We call this architecture CNN-Trans.  We evaluate it on AP17-OLR data and the MLS14 set of NIST LRE 2017, and show that the PHO-LID model with multi-task optimization exhibits the highest LID performance among all models, achieving over 40\% relative improvement in terms of average cost on AP17-OLR data compared to a CNN-Trans model optimized only for LID. The visualized confusion matrices imply that our proposed method achieves higher performance on languages of the same cluster in NIST LRE 2017 data than the CNN-Trans model. A comparison between predicted phoneme boundaries and corresponding audio spectrograms illustrates the leveraging of phoneme information for LID.
\end{abstract}
\noindent\textbf{Index Terms}: Language identification, acoustic phonetics, phonotactics, self-supervised learning, phoneme segmentation

\section{Introduction}
\label{sec:intro}
Spoken language identification~(LID) refers to the process through which the language identity of a speech sample can automatically be determined \cite{li2013spoken}. Early studies in LID suggested that acoustic and phonotactic features are the most effective language cues \cite{li2013spoken,comp4appr,reviewlid}. In particular, acoustic features such as Mel-frequency cepstral coefficients~(MFCCs) and filter bank energies are the most commonly used language cues in recent LID systems \cite{mfccs,dehak2011language,cai2019utterance}. Existing acoustic LID models, in general, comprise a language encoder and a classifier. They can either be optimized separately such as the conventional i-vector and x-vector methods \cite{dehak2011language, snyder2018spoken}, or integrated in an end-to-end neural network \cite{Miao2019,ling20_odyssey}. In addition, the phoneme-aware acoustic LID method utilizes phonemic information to achieve high LID performance. Such acoustic-phonetic LID system is trained on acoustic features and incorporates phoneme information by jointly optimizing the LID task and a phoneme-related task such as automatic speech recognition and phoneme classification \cite{ling20_odyssey,phoneawLID,phoneawSV}. In existing phoneme-aware LID methods, the LID and phoneme-related tasks share the same language encoding module in which the language identities and phoneme information are learned. As a result, the phoneme-related branch requires phoneme annotations or transcriptions.

Phonotactics involves the permissible combinations of phonemes in languages. As opposed to an acoustic-phonetic unit that is shorter than a phoneme, a phonotactic unit can cover several phonemes. It is also useful to note that while phonemes can be shared across languages, the statistics of their sequential patterns differ from one language to another \cite{li2013spoken}. Conventional phonotactic LID methods comprise a phone recognition module followed by the language modeling for target languages \cite{comp4appr,tong09_interspeech}. Typically, in the parallel phone recognition language modeling~(PPRLM) LID model \cite{comp4appr}, a phone recognizer that can either be universal or language-specific is first trained. The language models then capture sequential patterns of the recognized phones before generating scores for the target languages.
% Recent work explored the phonotactic information encoded in the self-supervised learning representation (wav2vec features) for the LID task \cite{ramesh21_interspeech,wav2vec}.

Notwithstanding the above, a phonotactic LID system faces the same challenge as the phoneme-aware LID method\textemdash it requires phoneme annotations of speech samples during training. However, the process of annotation is usually time consuming, requires significant resources and domain expertise. In contrast, acoustic approaches require only digitized speech samples and their language labels; they have since gained popularity over the recent years.

% In existing phoneme-aware LID methods, the LID and phoneme-related tasks share the same language encoding module through which the model learns language identities and phoneme information. The output of the encoder is then fed into the task-specific classifiers for multi-task learning \cite{phoneawLID,phoneawSV}. Consequently, they meet the same challenge as the phonotactic approach that the phoneme-related branch requires phoneme annotations or transcription.
% To achieve high LID performance, a straightforward approach is to combine multiple parallel LID subsystems with respect to different language cues via a back-end fusion of scores \cite{li2013spoken}. However, the subsystems and back-end fusion model are trained separately, which leads to a high complexity during training.

Since acoustic-phonetic and phonotactic cues depict the language identities of a speech signal in different granularities, it is natural to consider both jointly to achieve high LID performance. The proposed phonetic and phonotactic LID~(PHO-LID) method incorporates phonetic and phonotactic information hierarchically via a convolutional neural network-transformer encoder~(CNN-Trans) without the use of phoneme annotations \cite{transformer}. Inspired by the success of self-supervised phoneme segmentation in \cite{ssl_phoneme}, the CNN module in our model is shared by the LID and self-supervised phoneme segmentation tasks, through which it learns the language and positional phoneme information during training. To perform phonotactic LID, our proposed method mimics the phone n-gram modeling in a statistical manner. The language and phoneme-aware output features of the CNN module are first aggregated by a statistics pooling layer for each short-duration (hundreds of milliseconds) segment of the input speech signal. The segment-level phonotactic embeddings are then generated before the utterance-level LID, in which a statistics pooling layer computes the utterance-level statistics. 

Apart from the high LID performance resulting from the complementary characteristics of acoustic-phonetic and phonotactic features, our model possesses two desirable properties. Firstly, as opposed to existing phoneme or phonotactic-aware methods, our proposed approach does not require phoneme annotation or transcription due to the self-supervised training. Secondly, the proposed PHO-LID model combines different language cues in an end-to-end manner resulting in lower complexity compared to the fusion of several subsystems with respect to different language cues \cite{li2013spoken}.
\section{Related work}
\label{sec:related}
In \cite{ssl_phoneme}, the author proposed a self-supervised approach for phoneme segmentation. Given the representations of a raw speech signal, a CNN model is optimized to identify the spectral changes. This is achieved by minimizing a noise-contrastive estimation~(NCE) loss \cite{gutmann2010noise,oord2018representation}, in which the anchor, positive, and negative samples correspond to the current frame, its adjacent frames, and non-adjacent frames, respectively. 
% Each anchor frame is assumed to only belong to the same phoneme as its positive samples. 
Denoting $\mathrm{sim}(\cdot,\cdot)$ as cosine similarity between two vectors, the NCE loss for each frame \(\mathbf{z}_{i}\) is computed via
\begin{equation}
\setlength{\abovedisplayskip}{5pt}
\setlength{\belowdisplayskip}{5pt}
  \mathcal{L}\left (\mathbf{z}_{i}\right )=-\mathrm{log}\frac{e^{\mathrm{sim}\left ( \mathbf{z}_{i}, \mathbf{z}_{i+1} \right )}}{\sum_{\mathbf{z}_{j}\in \left \{\mathbf{z}_{i+1} \right \}\cup D_M\left ( \mathbf{z}_{i} \right ) }  e^{\mathrm{sim}\left ( \mathbf{z}_{i}, \mathbf{z}_{i+1} \right )} },
  \label{eq:loss_nce_frame}
\end{equation}
where \(D_M\left ( \mathbf{z}_{i} \right )\) is a set of $M$ negative samples to \(\mathbf{z}_{i}\). These negative samples are randomly selected from all non-adjacent frames. During inference, the similarities between every two adjacent frames are computed. The phoneme boundaries are then determined as where the similarity is lower than a predetermined threshold.
\section{Methodology}
\label{sec:method}
\subsection{PHO-LID model}
As shown in Fig.~\ref{fig:PHO}, the proposed PHO-LID model incorporates both acoustic-phonetic and phonotactic information hierarchically and comprises two branches. To illustrate these two branches separately, we define \(\mathbf{X} = (\mathbf{x}_{t}\in\mathbb{R}^{K\times F}| t=1, ... , T)\) as input of the proposed model. Here, \(\mathbf{X}\) comprises features extracted from the input speech signal partitioned in segments, and \(T\) is the number of segments. Each segment \(\mathbf{x}_{t}\) includes \(K\) frames \(\left [ \mathbf{f}_{t,1},...,\mathbf{f}_{t,K} \right ]^{\intercal}\), where \(\mathbf{f}_{t,k}\) is an \(F\)-dimensional feature vector of the \(k\)-th frame in segment \(t\), and the original speech signal consists of \(T\times K\) frames. It is useful to note that, in general, the average duration of common phonemes varies between 50~ms to 200~ms. The number of frames in each segment is thus defined to allow that each phonotactic unit in the proposed model can cover minimum two phonemes. These partitioned features are fed into the CNN module, which is jointly optimized by the primary LID task and the self-supervised phoneme segmentation task over each segment \(\mathbf{x}_{t}\).

\subsection{Self-supervised phoneme segmentation}
To extract phonotactic information, the model should assimilate phoneme information before capturing the sequential patterns across several frames. Similar to the method originally proposed in \cite{ssl_phoneme}, we employ the self-supervised phoneme segmentation to supply the CNN module with phoneme information.

\begin{figure}[t]
\setlength{\belowcaptionskip}{-0.2cm}
  \centering
  \includegraphics[width=\linewidth]{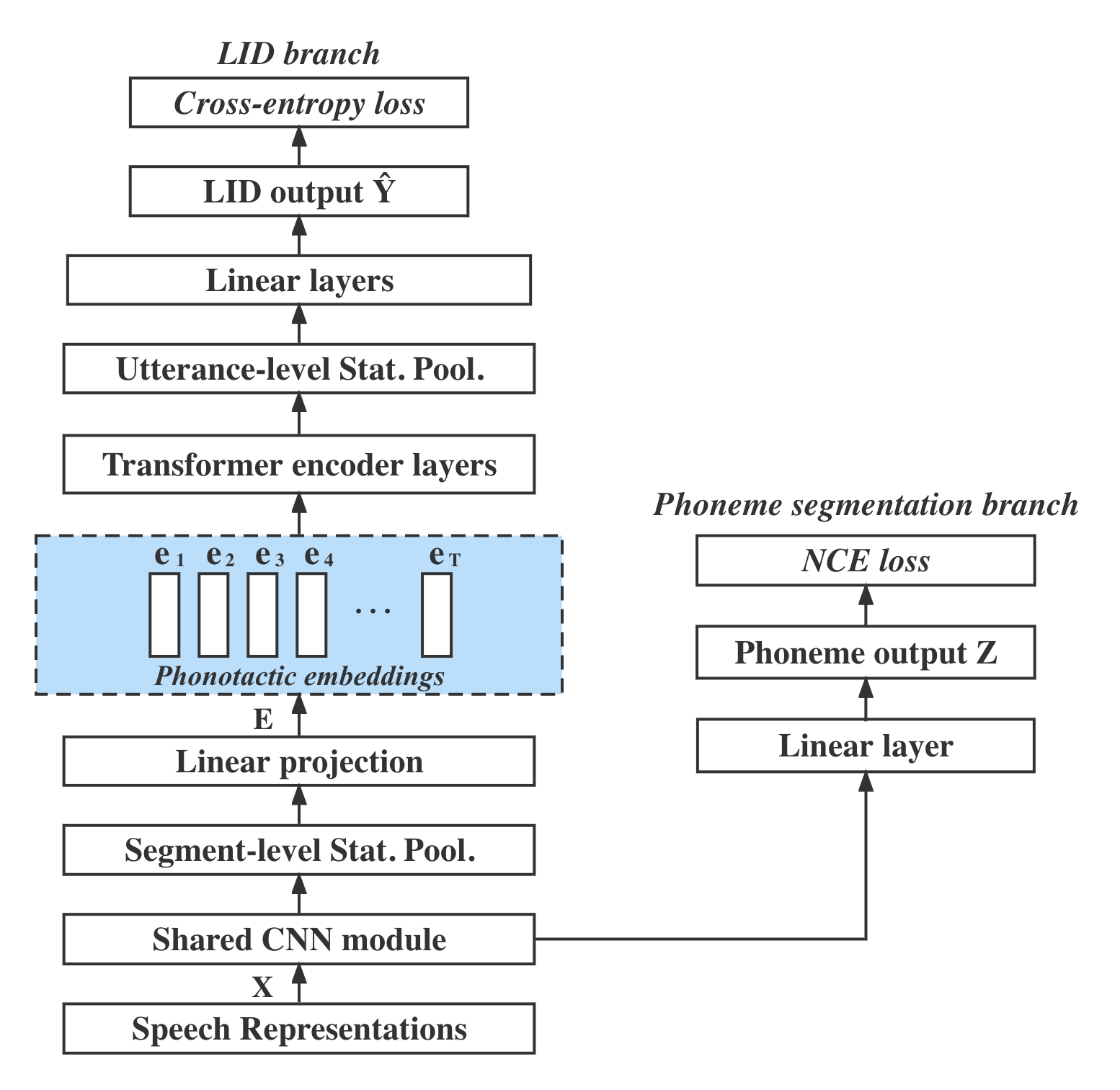}
  \caption{The PHO-LID model.}
  \label{fig:PHO}
\end{figure}

With reference to Fig.~\ref{fig:PHO}, the self-supervised phoneme segmentation branch consists of the shared CNN module and a linear layer. In \cite{ssl_phoneme}, given an input raw audio, each output unit of their model corresponds to a 10~ms frame, which is of a higher granularity than common phonemes. The premise that adjacents frames belong to the same phoneme is therefore reasonable. As opposed to \cite{ssl_phoneme}, the input features of our proposed model are of 25~ms receptive field with 20~ms step size similar to those in conventional LID methods. With a minimum phoneme length being approximately 50~ms, we adopt convolutional layers with a kernel size of 1 in the CNN module to achieve a shorter receptive field of the output unit than the minimum phoneme length. The output of the linear projection layer corresponding to the input segment-level features \(\mathbf{x}_{t}\) is given by \(\mathbf{Z} = (\mathbf{z}_{i} \in \mathbb{R}^{G}| i=1, ... , K)\), where $G$ is the output dimension of the phoneme segmentation branch. The utterance-level NCE loss is then computed via
\begin{equation}
\setlength{\abovedisplayskip}{5pt}
\setlength{\belowdisplayskip}{5pt}
  L^{NCE}=\frac{1}{KT}\sum_{T}\sum_{\mathbf{z}_{i}\in \mathbf{x}_{t}}^{K}\mathcal{L}\left (\mathbf{z}_{i}\right ),
  \label{eq:loss_nce_utt}
\end{equation}
where \(\mathcal{L}\left (\mathbf{z}_{i}\right )\) denotes the frame-level NCE loss defined in (\ref{eq:loss_nce_frame}).

As highlighted in Section~\ref{sec:related}, similarities between adjacent frames can reveal phoneme boundaries. Therefore, after optimization, positional phoneme information are intrinsically encoded in the phoneme segmentation branch. Since phonotactics characterizes the combinations of phonemes spanning a short proportion of speech, the learned phoneme information enables the proposed PHO-LID model to incorporate the phonotactic information in segment-level features.
\subsection{Supervised language identification}
As the CNN module learns the phoneme and language information for each segment \(\mathbf{x}_{t}\), the rest of the LID branch aims to perform phonotactic modeling. Features in each segment \(\mathbf{x}_{t}\) are then aggregated as mean and standard deviation vectors, which are concatenated into a representation vector. These representations are linearly projected to segment-level phonotactic embeddings \(\mathbf{E} = (\mathbf{e}_{t} \in \mathbb{R}^{D}| t=1, ... , T)\), with each $D$-dimensional vector \(\mathbf{e}_{t}\) corresponding to a segment \(\mathbf{x}_{t}\). This statistical process serves as the phone n-gram modeling of the input speech. The transformer encoder layers subsequently capture the global dependencies of these phonotatic embeddings \cite{transformer}, and the output is aggregated into an utterance-level language embedding. Consequently, a score vector $\widehat{\mathbf{Y}}$ corresponding to $C$ target languages is generated by the linear layers. 
% \subsection{Optimization strategies}

During training, we first pre-train the model on the self-supervised phoneme segmentation task for several epochs.Two strategies are employed and compared in the remaining updates. The first strategy solely updates the LID branch via a cross-entropy loss, while the second strategy optimizes the model by a multi-task objective function. These strategies are, respectively, given by
\begin{eqnarray}
\setlength{\abovedisplayskip}{5pt}
\setlength{\belowdisplayskip}{5pt}
  L^{LID} &=& \mathrm{CrossEntropy}\left ( \widehat{\mathbf{Y}}, \mathbf{Y} \right ),
  \label{eq:loss_lid_ce} 
  \\
  L^{MUL} &=& \alpha L^{LID} + \left ( 1-\alpha  \right ) L^{NCE},
  \label{eq:loss_lid_all}
\end{eqnarray}
where $\mathbf{Y}$ denotes the true language label and $\alpha$ is a parameter associated with multi-task learning. The model optimized by these two strategies are denoted as PHO-LID and PHO-LID-MUL, respectively. During inference, the input speech signal will be classified to the language of the highest score in $\widehat{\mathbf{Y}}$.

\section{Data and model configuration}
\label{sec:experiment}
\subsection{Datasets}
We evaluated our approaches on the AP17-OLR and NIST LRE 2017 dataset. The AP17-OLR dataset contains training, development, and test sets \cite{ap17olr}. The audios are of 16~kHz sampling rate and from ten languages. We trained our systems on the training set without THCHS30 and evaluated them on the test set. The NIST LRE 2017 dataset, on the other hand, consists of the Fisher corpus \cite{fisher}, Switchboard corpus \cite{switchboard}, a narrow-band telephony training set (TRN17) built from previous LRE data with over 2000 hours of audio data, a development set (DEV17), and an evaluation set (EVAL17) \cite{lre17_odyssey}. The DEV17 and EVAL17 comprise narrow-band MLS14 data and wide-band VAST data, while the MLS14 data consist of 3~s, 10~s, and 30~s duration and the VAST data comprise segments with speech duration ranging from 10 to 600~s. Fourteen languages exist in total within the data. We trained our systems on TRN17 and DEV17 and tested on the MLS14 data in EVAL17 to compare their performance on different duration levels. 

\subsection{Data preprocessing and feature extraction}
The recordings in TRN17 of the MLS14 data in NIST LRE 2017 are first upsampled from 8~kHz to 16~kHz before an energy-based voice activity detection~(VAD) is applied. The upsampled TRN17 recording are then partitioned into maximum 30~s segments before the feature extraction. The recordings in the AP17-OLR training set are partitioned into maximum 20~s segments without VAD before the feature extraction.

In this work, input features to the systems are wav2vec speech representations~(W2V) extracted from the 16th context network block of the XLSR-53 cross-lingual wav2vec 2.0 model \cite{wav2vec2,xlsr53}. This model has been pre-trained on fifty-three languages and 56,000 hours of speech data from Multilingual LibriSpeech, CommonVoice, and Babel \cite{MLS,commonvoice,babel}. The input wav2vec features $\mathbf{X}$ are of dimension $F=1024$ and partitioned into segments, each of which comprises $K=20$ frames before being fed into the model. Therefore, our phonotactic unit is approximately 400~ms long to cover minimum two phonemes. In addition, two baseline models in the experiments with respect to the NIST LRE 2017 data are trained on 80-dimensional bottleneck features~(BN). The bottleneck features are extracted from an ASR model pre-trained on the Fisher and Swithchboard corpora using the Kaldi toolkit \cite{KALDI}.

\subsection{Model configuration}
\label{sec:config}
We provide two baseline x-vector and XSA-LID models for the experiments of NIST LRE 2017. With configurations following that of \cite{snyder2018spoken} and \cite{liu2022enhance}, respectively. The x-vector is trained by modifying the SRE16 recipe in the Kaldi toolkit with a back-end logistic regression classifier. The XSA-LID model is trained following the same strategy as the CNN-Trans model.

In our proposed model, the shared CNN module comprises three convolutional layers with kernel sizes \(\left (1,1,1 \right )\) and output dimensions  \(\left (512,512,512 \right )\). The output dimension of the phoneme segmentation branch is given by $G=64$. The outputs of the segment-level statistics pooling layer are projected to $D$-dimensional phonotactic embeddings with $D=64$. The transformer encoder module consists of a layer norm followed by two transformer encoder layers, each of which has 8 heads, $d_{\mathrm{model}}=512$, and $d_{\mathrm{ff}}=2048$ \cite{transformer}. The following linear layers comprise 512, 512, and $C$ output nodes, respectively. The CNN-Trans model shares the same configuration as the PHO-LID model but excludes the phoneme segmentation branch.

The PHO-LID model is trained on the AP17-OLR data and NIST LRE 2017 data for 13 and 23 epochs, respectively. In the first 3 epochs, the CNN module is updated by the self-supervised phoneme segmentation task with a constant learning rate of \(10^{-4}\). In the remaining epochs, the model is updated with a learning rate that warms up from 0 to \(1\times 10^{-4}\) in 3 epochs followed by the cosine annealing decay. The PHO-LID-MUL employ two $\alpha$ values 0.95 for AP17-OLR data and 0.97 for NIST LRE 2017 data. The Adam optimizer with batch size 128 is used. We evaluated our systems by employing accuracy~(Acc.), equal error rate~(EER) and the average cost~(\({C}_{avg}\)) \cite{lre07}. The source code has been made available in GitHub.~\footnote{https://github.com/Lhx94As/PHO-LID}
\section{Experiment, results and analysis}
\begin{table}[t]
\centering
\caption{Evaluation on AP17-OLR by employing Accuracy~(\%), EER~(\%) and \({C}_{avg}\)}
\label{tab:ap17olr}
\setlength{\tabcolsep}{1.8mm}{
\begin{tabular}{|c|c|c|c|c|}
\hline
\textbf{Method} & \textbf{Feat.} & \textbf{Acc.}  & \textbf{EER}  & \textbf{\({C}_{avg}\)}         \\ \hline
i-vector \cite{dehak2011language,ap17olr} & MFCCs & - & 3.39 & 0.0352 \\ \hline
PTN-LID \cite{tang2017phonetic,ap17olr} & Fbanks & - & 8.15 & 0.0689  \\ \hline
CNN-Trans        & W2V  & 96.74   & 1.101           & 0.0098          \\ \hline
PHO-LID-3        & W2V  & 97.57   & 0.7972          & 0.0073 \\ \hline
PHO-LID-10       & W2V  & 97.67   & 0.8201          & 0.0075          \\ \hline 
PHO-LID-MUL-3    & W2V  & \textbf{98.13}   & \textbf{0.6303} & \textbf{0.0058} \\ \hline
PHO-LID-MUL-10   & W2V  & 97.50   & 0.8642          & 0.0080          \\ \hline
\end{tabular}}
\end{table}

\begin{table*}[t]
\centering
\caption{Evaluation on the MLS14 set of NIST LRE 2017 by employing Accuracy~(\%), EER~(\%) and \({C}_{avg}\) across three test speech durations 3~s, 10~s, and 30~s}
\label{tab:nistlre}
\setlength{\tabcolsep}{3.0mm}{
\begin{tabular}{|c|c|ccc|ccc|ccc|}
\hline
\multirow{2}{*}{\textbf{Method}} & \multirow{2}{*}{\textbf{Feat.}} & \multicolumn{3}{c|}{\textbf{3s}} & \multicolumn{3}{c|}{\textbf{10s}} & \multicolumn{3}{c|}{\textbf{30s}} \\ \cline{3-11} 
&   & \textbf{Acc} & \textbf{EER} & \textbf{\({C}_{avg}\)} & \textbf{Acc} & \textbf{EER} & \textbf{\({C}_{avg}\)} & \textbf{Acc} & \textbf{EER} & \textbf{\({C}_{avg}\)} \\ \hline
x-vector \cite{snyder2018spoken}                    & BN    & -      & 11.79  & 0.1159   & -      & 7.81    & 0.0748   & -       & 6.84   & 0.0654        \\ \hline
\multirow{2}{*}{XSA-LID \cite{liu2022enhance}}      & BN    & 54.18  & 15.47  & 0.1685   & 74.90  & 7.39    & 0.0739   & 84.09   & 4.46   & 0.0406        \\ \cline{2-11}
                                                    & W2V   & 70.16  & 10.71  & 0.1004   & 84.32  & 4.89    & 0.0432   & 87.20   & 3.71   & 0.0329        \\ \hline
CNN-Trans       & W2V  & \textbf{73.63} & \textbf{7.89} & 0.0710          & \textbf{86.36} & 3.89          & 0.0346         & 89.93          & 2.91           & 0.0262          \\ \hline
PHO-LID-3       & W2V  & 72.61          & 8.22          & 0.0712          & 85.89          & \textbf{3.84} & 0.0344         & \textbf{90.00} & \textbf{2.77}  & \textbf{0.0242} \\ \hline
PHO-LID-MUL-3   & W2V  & 72.52          & 8.10         & \textbf{0.0696} & 84.90          & 3.86          & \textbf{0.0332} & 89.81          & 2.87          & 0.0253          \\ \hline
\end{tabular}}
\end{table*}

\subsection{Performance on AP17-OLR data}
\label{sec:ap17olr}
We present the evaluation results of the baseline and our proposed models on AP17-OLR data in Table~\ref{tab:ap17olr}. Results of the i-vector and phonetic temporal neural~(PTN-LID) models are provided by the AP17-OLR evaluation plan \cite{dehak2011language,tang2017phonetic,ap17olr}. The averaged results of the models across several trials are presented in Table~\ref{tab:ap17olr}. Here, the suffixes of PHO-LID and PHO-LID-MUL denote the number of negative samples $M$ in (\ref{eq:loss_nce_frame}). In Table~\ref{tab:ap17olr}, the PHO-LID-MUL-3 model achieves the highest LID performance compared to other methods and exhibits 40.82\% relative improvement in terms of \({C}_{avg}\) compared to the CNN-Trans model. This indicates that incorporating acoustic-phonetic and phonotactic information effectively enhances the LID performance. In addition, the PHO-LID-MUL-3 model achieves higher performance in terms of EER and accuracy compared to the PHO-LID-3. This suggests that the performance improvement is mainly attributed to the pre-training for self-supervised phoneme segmentation.

As $M$ increases, the PHO-LID-MUL model suffers from significant performance degradation while model exhibits few variations of LID performance. Comparing the performance between the PHO-LID-MUL-3 and PHO-LID-3 models, these results imply that a small $M$ with multi-task learning is preferred.

\begin{figure}[t]
\setlength{\belowcaptionskip}{-0.2cm}
  \centering
  \includegraphics[width=\linewidth]{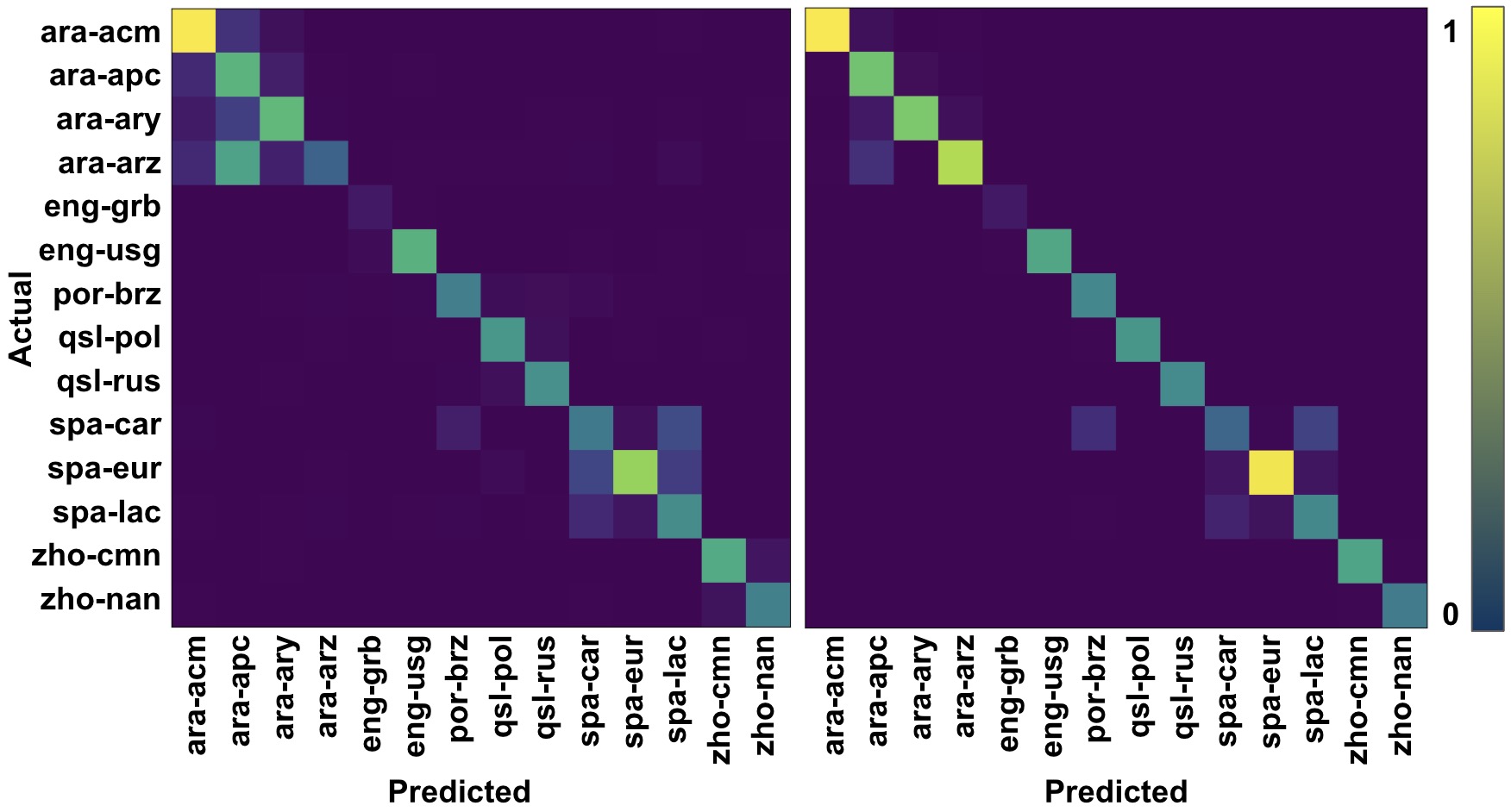}
  \caption{Performance confusion matrices of the CNN-Trans model (left) and the PHO-LID-MUL-3 (right) on 30~s test speech in NIST LRE 2017 data.}
  \label{fig:confusionmatrix}
\end{figure}
\subsection{Performance on the MLS14 set of NIST LRE 2017}
Table~\ref{tab:nistlre} presents the results on the MLS14 set of NIST LRE 2017 by the test speech durations. The proposed PHO-LID-MUL-3 model exhibits the highest overall performance compared to other systems in terms of \({C}_{avg}\), and both PHO-LID-3 and PHO-LID-MUL-3 models outperform the CNN-Trans model on overall performance in terms of \({C}_{avg}\). The above is consistent with the results shown in Table~\ref{tab:ap17olr} and suggests the effectiveness of our proposed method. Moreover, compared to the CNN-Trans model, while the PHO-LID-MUL-3 achieves 1.97\% relative improvement on 3~s test speech in terms of \({C}_{avg}\), it achieves 4.05\% and 3.44\% relative improvement on 10~s and 30~s speech, respectively. This is not surprising since phonotactic features are extracted from longer speech units as opposed to acoustic features.

We visualize the confusion matrices of the LID performance of the CNN-Trans and PHO-LID-MUL-3 models on 30~s test speech in Fig.~\ref{fig:confusionmatrix}. Compared to the CNN-Trans model, the proposed PHO-LID-MUL-3 model with phonemic and phonotactic information effectively reduces the confusion resulting from languages belonging to the same cluster\textemdash Arabic and Iberian clusters, and thus achieves higher performance.

In addition to the above, results presented in Tables~\ref{tab:ap17olr} and \ref{tab:nistlre} show that the W2V features introduce significant performance improvement compared to the use of conventional acoustic-phonetic features. This suggests that the W2V features are more discriminative and well-suited for the LID task.
\section{Visulization and discussion}
With reference to \cite{ssl_phoneme}, phoneme boundaries occur where the similarity of two adjacent frames is lower than a predefined threshold. To demonstrate the existence of the positional phoneme information in our proposed model, we compare similarities between adjacent frames in a speech sample with the corresponding spectrogram. Since phoneme boundaries detection is not our target during inference, we annotate regions with significantly low similarities to indicate the predicted phoneme boundaries using dashed lines. 

\begin{figure}[t]
\setlength{\belowcaptionskip}{-0.2cm}
  \centering
  \includegraphics[width=\linewidth]{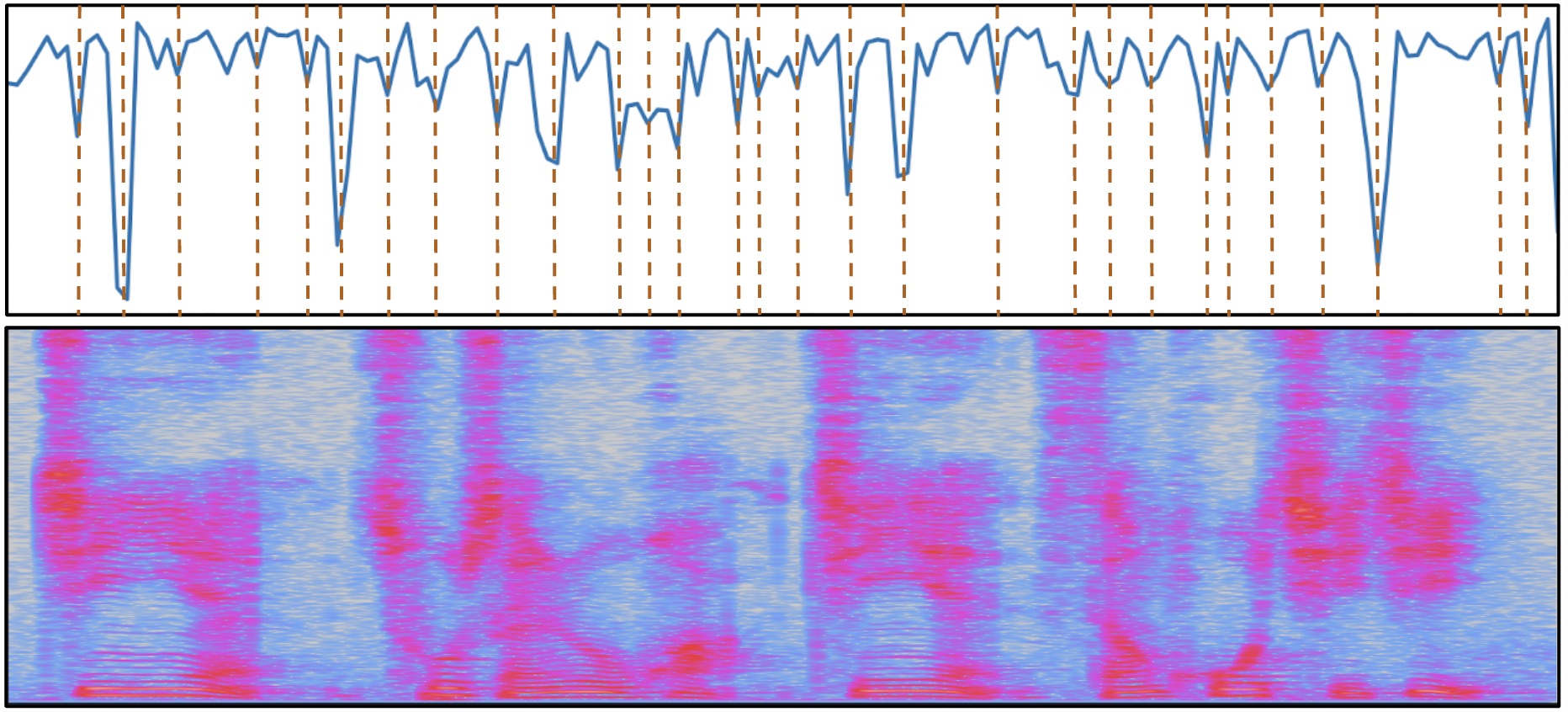}
  \caption{Comparison between the phoneme boundaries predicted by our model (top) and the audio spectrogram (bottom).}
  \label{fig:sim_spec}
\end{figure}

As shown in Fig.~\ref{fig:sim_spec}, the spectral changes can be approximately detected by our proposed model. This indicates that the model output contains positional phonetic information. Since our proposed phonotactic embedding is the concatenation of the mean and standard deviation vectors computed over frames in a segment, the above comparison highlights the feasibility of incorporating phonetic and phonotactic information. 

\section{Conclusion}
We proposed the PHO-LID model that incorporates the phonetic and phonotactic information for LID without the need for phoneme annotations. Compared to the CNN-Trans model, our proposed model with multi-task optimization achieves higher performance on AP17-OLR and NIST LRE 2017 data, and the performance confusion matrices indicate that our proposed method can effectively distinguish languages of the same cluster in NIST LRE 2017. We visualize the predicted phoneme boundaries using the output units of the shared CNN module. The comparison between the predicted phoneme boundaries and the corresponding audio spectrogram shows the existence of phoneme information, which, in turn, highlights the feasibility of our proposed method.

\section{Acknowledgements}

This work was partially supported by the National Research Foundation, Singapore, under the Science of Learning programme (NRF2016-SOL002-011), the Centre for Research and Development in Learning (CRADLE) at Nanyang Technological University, and US the National Science Foundation via CCRI Award \#2120435.

\newpage

\bibliographystyle{IEEEtran}

\bibliography{mybib}

\end{document}